\documentclass[aps,groupedaddress,superscriptaddress,amsmath,amssymb,prb]{revtex4-2}

\usepackage{bm}
\usepackage[pdftex]{graphicx}
\usepackage{xcolor}
\usepackage{color}
\usepackage[normalem]{ulem}
\usepackage{enumerate}
\usepackage{epstopdf}
\usepackage{braket}
\usepackage{hyperref}
\usepackage{blindtext}
\usepackage{comment}
\usepackage{amsmath}
\usepackage{float}

\renewcommand{\figurename}{Figure}

\begin{document}

\title{Cascade of multi-exciton states generated by singlet fission }

\author{Yan Sun}
\affiliation{LPS, Universit\'e Paris-Saclay, CNRS, UMR 8502, F-91405 Orsay, France}
\author{M. Monteverde}
\affiliation{LPS, Universit\'e Paris-Saclay, CNRS, UMR 8502, F-91405 Orsay, France}
\author{V. Derkach}
\affiliation{O. Ya. Usikov Institute for Radiophysics and Electronics of NAS of Ukraine 12, Acad. Proskury st., Kharkov, 61085, Ukraine }
\affiliation{LPS, Universit\'e Paris-Saclay, CNRS, UMR 8502, F-91405 Orsay, France}
\author{J. E. Anthony}
\affiliation{Department of Chemistry, University of Kentucky, Lexington, KY 40506-0055, USA }
\author{A.D. Chepelianskii}
\affiliation{LPS, Universit\'e Paris-Saclay, CNRS, UMR 8502, F-91405 Orsay, France}

\begin{abstract}
Identifying multi-exciton states generated from singlet fission is key to understanding the carrier multiplication process, which presents a strategy for improving the efficiency of photovoltaics and bio-imaging. Broadband optically detected magnetic resonance (ODMR) is a sensitive technique to detect multi-exciton states. Here we report a dominant species emerging under intense light excitation corresponding to a weakly exchange coupled triplet pair located on adjacent molecules oriented by nearly 90-degree ($V2$), contrasting to the $\pi -$stacked triplet pair under low excitation intensity. The weakly coupled species model precisely reproduces the intricate spin transitions in the Hilbert space of the triplet pair. Combining the magneto photoluminescence and high-magnetic field ODMR, we also identify a strongly exchange-coupled state of three triplet excitons formed by photoexcited $V2$, which manifests through the magnetic field induced level crossings between its quintet and triplet manifolds. The excellent agreement between the experimental Zeeman fan and the two-triplet spin Hamiltonian highlights the potential of multi-exciton states for quantum information processing. 
\end{abstract}


\maketitle

Singlet fission represents a distinctive carrier multiplication phenomenon observed in certain organic materials, wherein a singlet exciton undergoes decay, generating two lower-energy triplet excitons, each possessing approximately half the energy of the initial singlet state. Recent investigations have showcased the potential of this process in significantly enhancing the efficiency of solar cells  \cite{Hanna2006,tayebjee2012,Ehrler2012,lee2013,Rao2017}, pushing them beyond the confines of the Shockley-Queisser limit  \cite{Congreve2013,Dirk2018,Einzinger2019,Tayebjee2022,Hudson2022}. The prevailing understanding posits the formation of a transient bi-exciton during singlet fission, serving as a mediator between the photoexcited singlet and two dissociated triplet excited states  \cite{Smith2010,Swenberg1968,kazzaz1968}. Likewise, a comparable transient triplet pair is thought to function as an intermediate state in the reverse process of triplet-triplet annihilation. This reverse mechanism culminates in upconversion, where two low-energy triplet excitons coalesce into a higher-energy singlet excitation  \cite{Keivanidis2003,Singh2010,Schmidt2010,Pandey2015,Wenping2021,Bossanyi2021}. The implications of such upconversion extend to potential applications in catalysis  \cite{Ravetz2019,Castellano2011}, bio-imaging \cite{Liu2012,Liu2018} and high precision volumetric printing \cite{sanders2022}.

While bi-excitons states have been identified in various materials \cite{Bryant1990,PhysRevLett.64.1805,PhysRevLett.88.117901,Baldo2000,Klimov1998,Masui2013}, characterizing these states optically is often challenging. Theoretical predictions of their spectroscopic properties prove delicate, and in organic materials, their spectra often overlap with those of singly excited states, preventing their reliable identification through traditional spectroscopic methods. For singlet fission time-resolved photoemission spectroscopy \cite{Chan2011} and optical pump-probe spectroscopy have allowed the observation of intermediate states, hinting at ultra-fast separation dynamics and the existence of potential multi-exciton transient intermediates \cite{Wilson2011,Pensack2016,Bakulin2016,Monahan2017}. Spin resonance methods, capable of resolving higher spin states arising from spin-spin coupling \cite{teki2000intramolecular,teki2001pi,teki2008design,bayliss2015spin}, have been instrumental in establishing the existence of bound bi-exciton states in the context of singlet fission, providing compelling evidence of states with total spin S=2, often referred to as quintets in spin resonance literature \cite{Tayebjee2017,Weiss2017}. Fluorescence spectroscopy at high magnetic field experiments confirmed the level anti-crossings between singlet S=0, triplet S=1, and quintet S=2 allowing to quantify the strong exchange interaction between the two paired triplet excitons \cite{bayliss2018,bayliss2016spin,wakasa2015can,chakraborty2014massive,huang2021competition,yago2022triplet}.

Highly sensitive optically detected magnetic resonance (ODMR) \cite{shinar2012optically,clarke1982triplet,schmidt1968optical,wrachtrup1993optical,kohler1993magnetic,laplane2016}, enables the broadband measurement of the spin energy spectrum's dependence on the magnetic field allowing to identify the underlying spin-Hamiltonian with dipole-dipole interactions providing a precise identification of microscopic nature of the bound bi-exciton state. In TIPS-tetracene, a reference system with efficient singlet fission \cite{stern2015identification,stern2017vibronically}, this method showed that the two bound triplet excitons in this system reside on adjacent $\pi$-stacked TIPS-tetracene dimers \cite{ODMR2019,sun2023spin}. The excellent agreement between structural parameters obtained from spin resonance and crystallographic data demonstrated that bi-excitons are intrinsic to the pristine system and not induced by defects. This approach has been adopted to investigate the bi-exciton structure in a newly synthesized fission material \cite{Gajadhar} and to investigate the fundamental mechanisms driving the spin-dependent recombination of quintet bi-excitations \cite{sun2023spin}.

Here, we investigate multi-exciton states emerging at higher light intensities aiming to understand their microscopic structure, and to find possible evidence for long-lived multi-exciton states with more than two excitons. For applications, it is also important to understand these species to control exciton-exciton annihilation effects which can limit exciton fission yield. We focus on the fission material TIPS-tetracene, building on the well-established understanding of the properties of the geminate quintet state. We show that at high excitation intensities, broadband ODMR reveals a new species with a very complex pattern of magnetic resonance transitions. We model these spin transitions as a triplet pair with excellent precision observing transitions between all possible multiplicities (quintet S=2, triplet S=1, and singlet  S=0). This enables precise identification of its structure as a weakly coupled pair of triplet excitons on the nearest molecules oriented by around 90-degree($V2$).  
In addition to this weakly coupled species, high-field level crossing experiments demonstrate a growing level crossing feature at 5.5 Tesla which might be a signature of a higher-order spin state. To resolve the microscopic structure of this state we expanded the frequency range of our ODMR setup to enable experiments directly at the high magnetic field (5.5 Tesla) level crossing. We also measured the fluorescence spectrum associated with particular ODMR transitions allowing us to compare it with the magnetic field-dependent fluorescence spectrum associated with the level crossings. This allows us to establish a deep connection between the strongly coupled state and the weakly coupled exciton pair.

\begin{figure}
\includegraphics[clip=true,width=0.7\columnwidth]{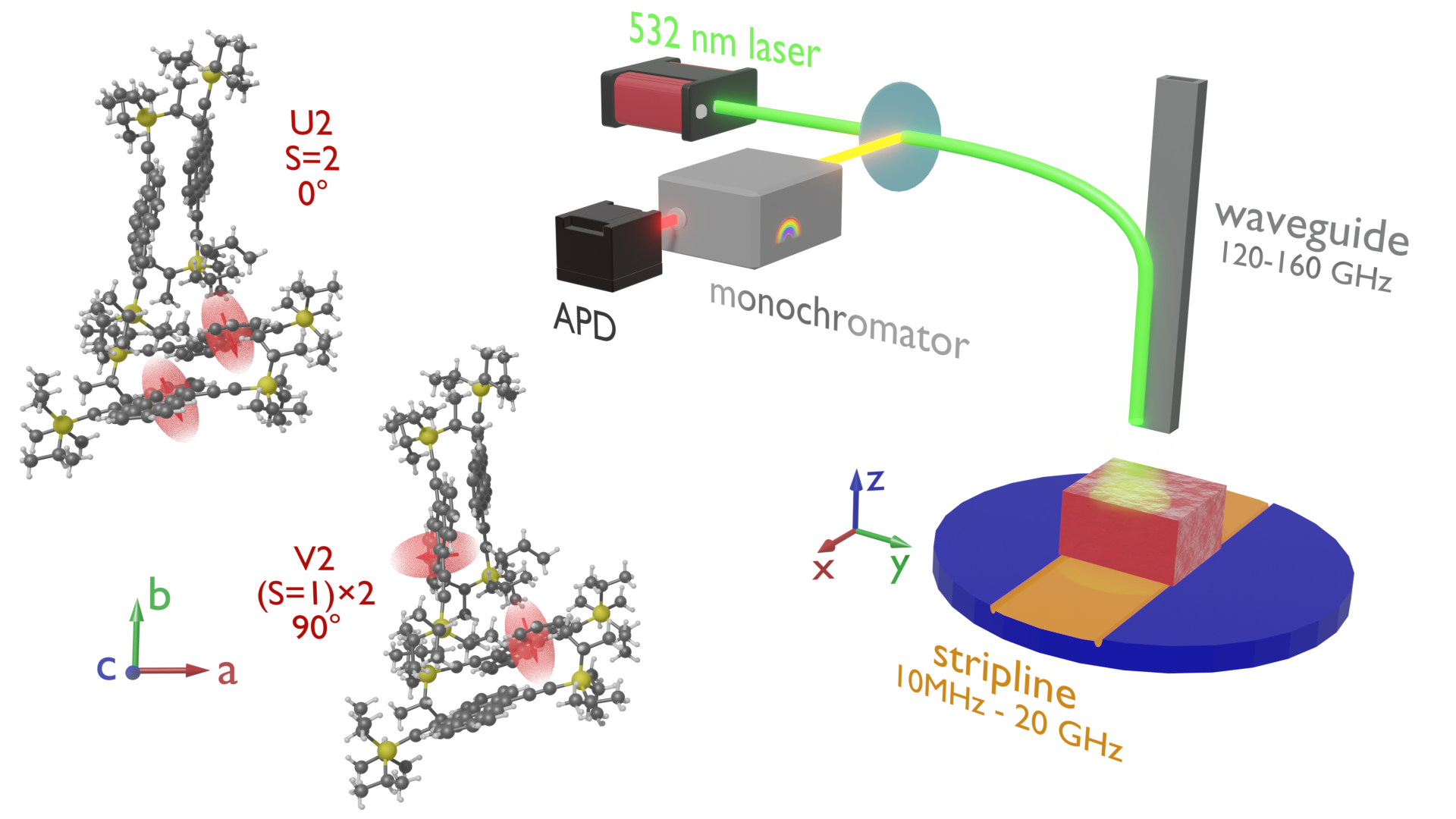}    
\caption{Sketch of the setup. The crystalline sample is placed on the stripline on the sample holder, which is closely installed underneath the fiber and waveguide. The excitation light of the 532 nm laser and the fluorescence from the sample are transmitted through the fiber. Depending on the magnetic field range, the microwave frequency is sent through the stripline or waveguide.  An APD is used to detect the fluorescence signal at the microwave on/off modulation frequency for ODMR. It can be placed after a monochromator to measure the change in the fluorescence spectrum induced by a specific magnetic resonance transition. The left insert shows the crystal cell of TIPS-tetracene unit cell with four molecules and two multi-exciton states that we reliably identify through ODMR. The $\pi-$ stacked quintet state (U2) is shown on the top and is the dominant long-lived states at low light intensity. The bottom structure shows the 90-deg bi-exciton (V2) which we show to become the dominant long-lived state at high excitation.}
\label{figSample}
\end{figure}

A sketch of our setup is shown in Fig.~\ref{figSample}, a deuterated TIPS-tetracene crystal is placed on a microwave strip-line and close to the end of a waveguide allowing to excite the sample at 150GHz frequencies which is the frequency range of the Zeeman splitting at the 5.5 T level-crossing. Deuterated samples were used as in \cite{sun2023spin} because they provide a sharper magnetic resonance transition. Light excitation and fluorescence collection are ensured by a multimode fiber placed close to the samples termalized at 4.2K in helium vapor. The fluorescence is then sent on an avalanche photodiode (APD) where the ODMR signal can be detected at the amplitude modulation frequency of the microwave signal in the strip-line or of the mm-wave waveguide excitation depending on the magnetic field range. For wavelength resolved ODMR, the APD is placed after a monochromator which can select a specific wavelength for the PL detection. For magneto-fluorescence experiments, the total (non-modulated) PL signal is detected. The 532nm laser excitation is amplitude stabilized through a PID loop which also controls the excitation intensity.

\begin{figure}
\includegraphics[clip=true,width=0.95\columnwidth]{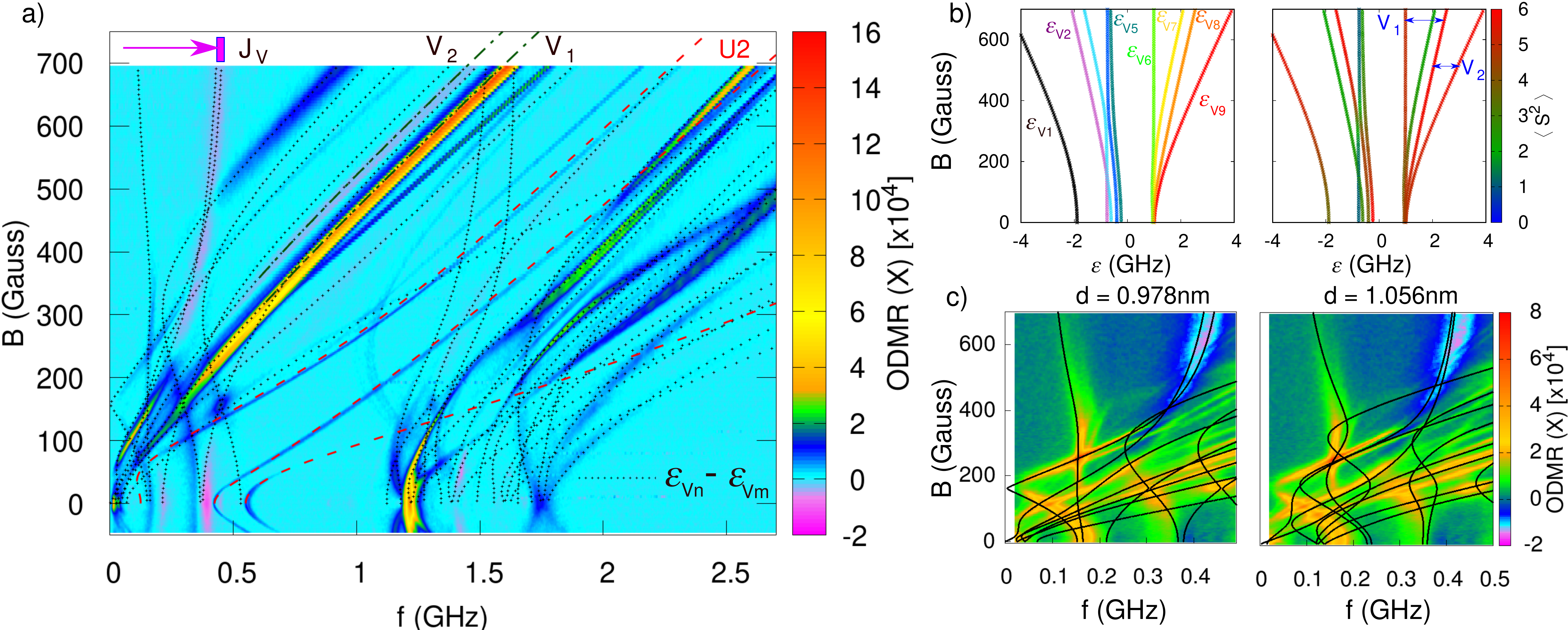}  
\caption{ODMR map shows the coexistence of two bi-exciton states with molecular structure shown in Fig. 1. The transitions shown by dotted lines correspond to the weak-coupled triplet-pair ($V2$), precisely reproducing most of the visible experimental transitions (some of the predicted transitions which are missing become visible at higher as shown in supplementary information). The theoretical energy level diagram giving rise to this pattern is represented in panel b) which also shows the average spin $\langle S^2 \rangle$ for each of the energy-eigenstates of the triplet pair. This allows us to identify states which have mainly quintet characters ($\langle S^2 \rangle \simeq 6$) giving rise to transitions $V_{2,1}$. The vertical transition marked as $J_V$ corresponds to transitions between states of the same angular momentum from the quintet and triplet manifolds. Its position gives the exchange coupling (see main text). Three transitions from the $U2$ are highlighted by red-doted lines. In c) the low-frequency transition is measured more precisely with a microwave-power amplifier. Simulated transition lines from weakly coupled triplet pair for two possible geometries are compared with the experiment. The nearest 90-deg stacked pairs with d = 0.978 nm (left) show a better overlap compared to the next neighbor 90-deg sites with d = 1.056 nm (right). }
\label{figmap}
\end{figure}

A frequency/magnetic field ODMR map at high light intensity is shown in (Fig.~\ref{figmap}) revealing a multitude of spin transitions.
In the ODMR map, a few transition lines from  $\pi -$stacked quintet U2 are visible under a low magnetic field, but their relative intensities are weaker and drop with increasing magnetic field. This is due to the quick saturation of the ODMR signal from U2 with the light intensity. The dominant transitions of the Zeeman-fan diagram have a non geminate character disapearing at low light intensity (see S.I.). They can be explained by a weakly coupled triplet pair model, its spin Hamiltonian is given by:
\begin{align}
  \hat{H_V}=\sum_{\nu=a,b}{\hat S}_{\nu}^{T}\cdot\mathbf{D}_{T}^{\nu}\cdot {\hat S}_{\nu}-\Delta(\mathbf{u}_{ab}\cdot\mathbf{{\hat S}}_{a})(\mathbf{u}_{ab}\cdot\mathbf{{\hat S}}_{b})-J_V \mathbf{{\hat S}}_{a}\cdot\mathbf{{\hat S}}_{b}+\sum_{\nu=a,b}\mathbf{\hat S}_{\nu}\cdot{\hat B}
  \label{Htt}
\end{align}
where $a$ and $b$ are two sites of occupied by the pair of triplet excitons which are taken from the crystal structure. The fine structure tensor $\mathbf{D}_{T}^{\nu}$ is determined by the structure parameter ${D}_{T}= 1.4$ GHz and is oriented along the direction of the $\pi$ orbitals. The dipole-dipole interaction depends on the distance between the two molecules in which the excitons are located. Their strength is given by $\Delta = \frac{3\mu_{0}\mu_{B}^{2}g^{2}}{4\pi r_{ab}^{3}}$, and orientation is given by $\mathbf{u}_{ab}$. The exchange interaction ${J}_V$, is the main unknown parameter in the model. The magnetic field orientation with respect to the cristal is fixed precisely by the U2 transitions.  We first attempted an extensive numerical optimization of the Hamiltonian parameters to reproduce the experimental lines using our procedure for \cite{sun2023spin}. However, this approach failed to converge to a parameter set reliably reproducing the experimental transitions. The key to the interpretation of the spectrum was to realize that the exchange coupling $J_V$ can be directly read from the ODMR map fixing precisely one of the parameters. Its value is fixed by the position of the vertical transition highlighted in Fig.~\ref{figmap}. Such vertical lines are unusual for magnetic resonance as we expect the resonance frequencies to grow with the magnetic field due to increasing Zeeman splitting. However, transition energies between spin states with the same angular momentum projection quantum number will not increase with the magnetic field due to cancellation between the Zeeman energies. For a triplet pair the wavefunction of the +1 state is $\frac{1}{\sqrt{2}} \left( |1,0\rangle -  |0,1\rangle \right)$, it differs only in the $'-'$ sign of the superposition from the +1 state of the quintet pair: $\frac{1}{\sqrt{2}} \left( |1,0\rangle + 0,1\rangle \right)$. Thus the energy difference between these two states will only depend on the interaction terms between the two triplets. This allows us to fix $J_V = 250MHz$ (as this exchange interaction is still large compared to dipole-dipole interaction) obtaining a very good agreement with the predictions with the Hamiltonian $H_V$ (Eq.1). The ODMR sign for this transition is negative contrarily to most other $V2$ transitions. This drop in PL is consistent with the interpretation that the bi-exciton $V2$ state has a higher triplet anhilation rate in its triplet manifold compared to its quintet manifold \cite{ha2022exchange} leading to a drop in PL under microwave excitation.

In the panels Fig.~\ref{figmap}.b we show the theoretical eigenvalue spectrum of $V2$, as well as the average of the total spin operator for each eigenstate (Fig.~\ref{figmap}.c). While at low magnetic field there is a mixing between different multiplicities, the eigenstate separates into a well-defined quintet, triplet and singlet states at high field, which can be seen from the $\langle S^2 \rangle$ which is close to $6, 2$ and $0$ for quintet, triplet and singlet respectively. With small exchange coupling($J_V < D_T$), well-defined quintet states are visible, which is possible because of the same angle between the magnetic field and the fine structure tensor for both triplets. Demonstrating the existence of weakly coupled quintets is useful for the interpretation of Rabi-oscillation experiments showing that S=2 states do not necessarily require a strong exchange coupling ($J_V \gg D_T$). The detailed agreement between the experimental map and the theoretical spectrum is confirmed by measuring the ODMR signal with a microwave power amplifier which reveals some hidden transitions but at the cost of introducing spurious harmonics in the signal (see S.I). 

The four nonequivalent molecules in the unit crystal cell of TIPS-Tetracene yield several possible configurations for (approximately) 90-degree bi-excitons. The exact microscopic position of the pair can be identified by the fine structure of the low-frequency transitions, which is very sensitive to the positions of triplet excitons due to changes in the tilts between molecules and in dipolar interaction. 
Fig.\ref{figmap}d shows the difference between the nearest neighbors pair with $d=0.978\ nm$ and the next neighbor with $d = 1.056\ nm$. 
As the nearest neighbor pair provides a better agreement with the experiment, it is reasonable to identify the $d = 0.978\ nm$ pair as the dominant geometry for the $V2$. Remarkably the exchange energy of $V2$ is orders of magnitude smaller than the $U 2$, revealing that the exchange energy between triplet excitons is extremely sensitive to the molecular geometry \cite{Taffet2020}. 

\begin{figure}
  \includegraphics[clip=true,width=0.8\columnwidth]{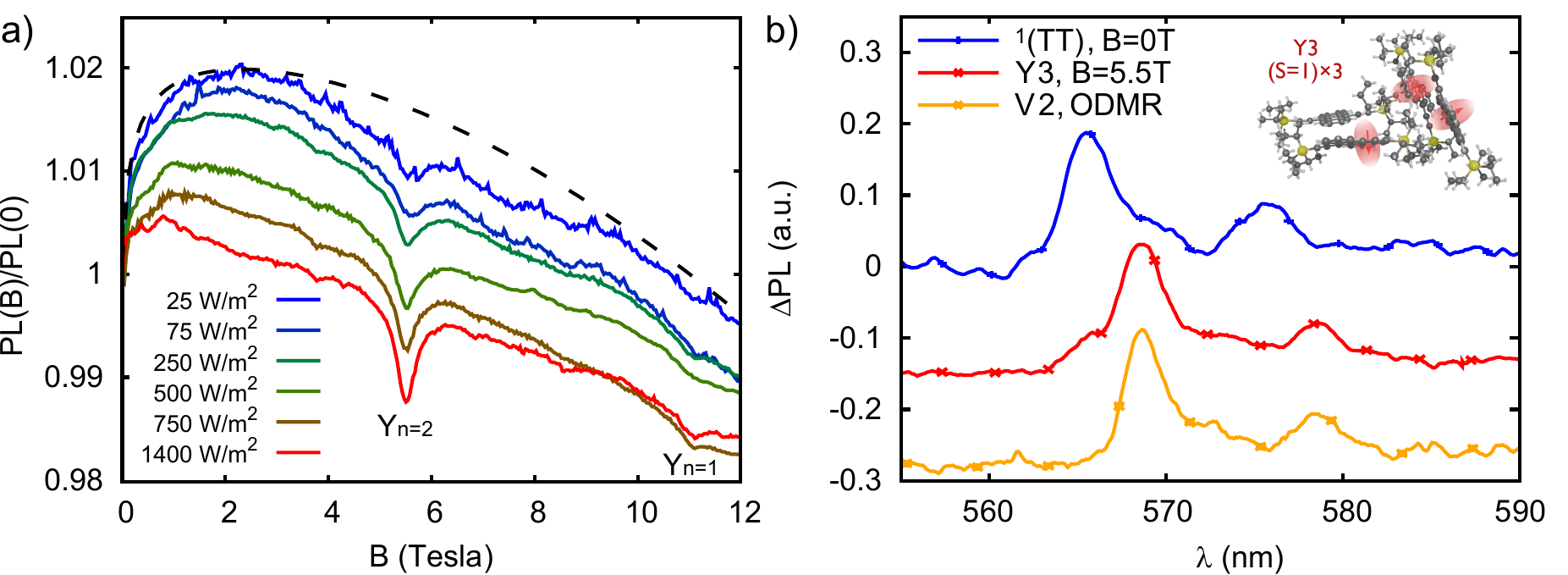}  
\caption{a) Evolution of the magneto-photoluminescence response with increasing excitation power showing the enhancement of the series of spectral holes $Y_{n=1,2}$ while the feature from zero-field weakens. The dotted line shows a simplified model of intersystem-crossing of a hot pair of triplets excitons $^1(TT)$ from singlet to triplet total spin state reproducing the low-fluency MPL. b) The photoluminescence spectra correspond to $^1(TT)$ at zero field (blue), $Y3$ the specie giving the $Y_{n=2}$ dip at 5.5 Tesla (red), and the weakly coupled 90-degree triplet pair $V2$ (orange). The match between the spectra from $Y3$ and $V2$ suggests a link between the two states. We propose that $Y3$ is a photo-excited $V2$ that decays into a three-exciton state (shown on the sketch).}
\label{figMPLandSpec}
\end{figure}

Magneto photoluminescence (MPL) experiments show two trends as the light intensity increases. At low light intensity, there is a strong low field MPL signal with an few \% increase of fluorescence at around 1 Tesla magnetic fields. A simple model inspired from \cite{merrifield} and reproducing the low-power MPL is presented in S.I., the result of the fit is also shown in Fig.~\ref{figMPLandSpec}a. The zero field feature is attributed to a short-lived hot singlet state, $^1(TT)$ \cite{sanders2019,carrod2022}, with lifetimes of around 200ps, in which a pair of triple-excitons make multiple hops between neighboring sites with low and high exchange energy (e.g. $V 2$ and $U 2$). This allows intersystem crossing of the triplet-pair reducing PL through the process $^1(TT) \rightarrow T^*$, where $T^*$ is an excited triplet state \cite{Ha2022}. To explain intersystem-crossing anti-symmetric exchange (Dzyaloshinskii-Moriya) is needed as all the matrix elements of the form $\langle T_{1,0,-1}| {\hat H}_V | S \rangle$  vanish for spin-Hamiltonians of the form Eq.~(\ref{Htt}). We estimate the strength of this interaction to around $20{\rm mTesla}$ (see S.I.). With the magnetic field rising, the Zeeman splitting overcomes the singlet-triplet intersystem crossing, resulting in increasing PL. The spin-allowed pathway $^1(TT) + T \rightarrow V2$ can lead to a quenching of $^1(TT)$ by triplet excitons, especially under a high concentration of excited $^1(TT)$ and long-lived triplet states. This is consistent with the drop of the low field MPL amplitude and V2 ODMR signal at low laser intensity.

In MPL (Fig.~\ref{figMPLandSpec}a), a growth of a strong-exchange coupled state at high light intensity is evidenced by 5.5 T level crossings ($Y3$). This crossing belongs to the same series as the less visible peaks at $11.0$ Tesla and $3.7$ Tesla, forming a $\Delta/n$ sequence ($n=1,2,3$ and $\Delta = 11$ Tesla). This series was reported previously in \cite{bayliss2018} for no-deuterated samples, along with another family with $\Delta' = 8.63 $ Tesla which is less visible in deuterated samples. The two families of crossings were attributed to quintets formed by triplet pairs located on different molecular sites. Here we identified the ODMR signatures of the two closest $\pi$-stacked ($U2$) and 90-degree stacked ($V2$) biexciton pairs. However, neither of them can explain the growing feature of high-field level crossing with increasing excitation power. Essentially the relative ODMR signal from $U2$ decreases with light intensity, while $V2$ doesn't have any high field level crossings due to its low exchange energy. As reported in \cite{bayliss2018}, the position of the PL dip gives a measure of the exchange energy of the triplet pair. It thus seems that there are no remaining candidate nearest neighbor pair sites for a strongly coupled triplet-pair state. This leads us to consider multi-exciton states with more than two triplet excitons. The interaction $^1(TT) + T$ which we invoked to explain the formation of $V2$ involves three interacting triplet excitons. However total spin for the three excitons participating in $^1(TT) + T$ is S=1, and there is no spin-selection rule preventing the formation of $T^*$ or triplet $V2$. It is thus difficult to explain a negative MPL peak with the $^1(TT) + T$ interaction as it already corresponds to the fastest possible non-radiative recombination. We can instead consider the state $Y3$ formed by a triplet exciton interacting with a triplet pair on the nearest $\pi$ and 90-degree stacked neighbor molecules (the insert in Fig.~\ref{figMPLandSpec}b). This state can be viewed as a photo-excited $V2$, as one of the two triplets formed by singlet fission can annihilate generating $T^*$ which keeps spin conserving in the manifold $S=2$. The state $Y3$ can also be formed through $U2 + T$ interactions at lower light intensity. We expect quintet $Y3$ to be as long-lived as $U2$ and triplet $V2$, which is sufficiently long-lived to give narrow transitions in magnetic resonance (see S.I. for a full map of the transitions). Under magnetic field, level crossing between the quintet and triplet manifolds of $Y3$ can lead to a dip in PL through the formation of $T^*$. In Fig.~\ref{figMPLandSpec}b, we show that the $\Delta$PL spectra associated with the MPL dip at 5.5 Tesla match the wavelength-resolved ODMR spectrum from $V2$. The last spectrum is obtained by filtering the PL through a grating spectrometer before the ODMR detection. The identical $\Delta$PL spectra from $V2$ and $Y3$ (5.5 Tesla PL dip) confirm the interconversion between the two species. On the other hand, the $\Delta$PL spectrum at $B=0$ is blue-shifted confirming its assignment to a hot state $^1(TT)$.

\begin{figure}
\includegraphics[clip=true,width=0.8\columnwidth]{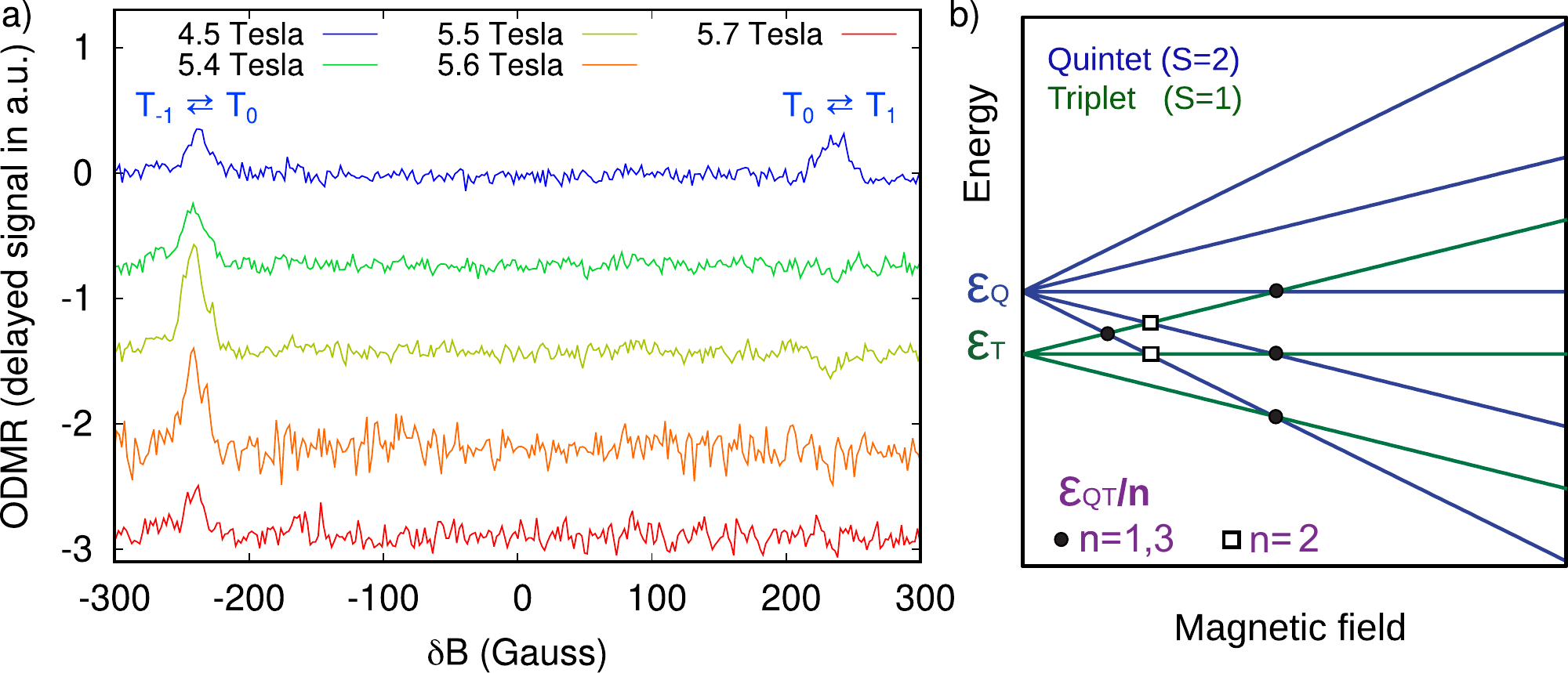}  
\caption{High field delayed ODMR spectra monitoring the transitions between triplets away and near the PL dip at 5.5 Tesla. While the transition amplitudes for  $T_0 \leftrightarrow T_{-1}$ and  $T_0 \leftrightarrow T_{1}$ are identical at $4.5$ Tesla, a strong asymmetry develops near $5.5$ Tesla together with an amplification of $T_0 \leftrightarrow T_{1}$. 
b) The asymmetry between $T_{1}$ and $T_{-1}$ can be understood from the level crossing energy diagram between quintet and triplet manifolds. At 5.5 Tesla both $T_0$ and $T_{1}$ intersect the quintet state. This can equilibrate their optical contrast leading to a vanishing ODMR signal while increasing the optical contrast for $T_0 \leftrightarrow T_{-1}$, providing evidence of an intersection between quintet and triplet manifolds.
}
\label{figmmODMRy}
\end{figure}

To confirm that the 5.5 Tesla MPL is due to the level crossing with a triplet state we performed ODMR experiments at this magnetic field. This requires an increase in the excitation frequency from a few GHz up to around  $150 {\rm GHz}$. In this frequency range, we have to use the waveguide for excitation. The triplet transitions preferentially show on the delayed channel of the ODMR signal due to the long triplet lifetimes \cite{bayliss2014geminate} (the delay is relative to the 537 Hz microwave power amplitude modulation). The delayed ODMR spectrum at several magnetic fields away and near 5.5 Tesla is shown in Fig.~\ref{figmmODMRy}.a. 
To compare the magnetic resonance lineshapes at different frequencies and $B$ we show them as a function of the magnetic field shifts $\delta B$ from the center of the ODMR pattern, in which the center field is indicated in legend. Away from the level crossing at $B = 4.5$ Tesla we see a similar signal from $T_0 \leftrightarrow T_{-1}$ and  $T_0 \leftrightarrow T_{1}$ transitions. This shows that ground state spin polarization is not an important source of asymmetry between  $T_0 \leftrightarrow T_{-1}$ and  $T_0 \leftrightarrow T_{1}$  most likely due to light-induced heating of the spins above the cryostat temperature of 4.2K.
However, near the level crossing the signal from $T_0 \leftrightarrow T_{1}$ substantially drops and even changes sign indicating a change in kinetics. On the other hand the  $T_0 \leftrightarrow T_{-1}$ transition seems amplified at $5.5$ Tesla (in SI we discuss the details of comparing signal at different mmwave frequencies). We can understand this behavior from the level crossing diagram in Fig.~\ref{figmmODMRy}.b. At the level crossing both $T_0$ and $T_1$ states intersect with the quintet manifold, this can lead to almost identical optical contrast between $T_0$ and $T_1$ or to an equilibration of the populations explaining the drop in ODMR signal. The optical contrast for $T_{-1} \leftrightarrow T_{0}$ can on the other hand increase as only $T_0$ crosses the quintet manifold. The analysis of the in-phase ODMR signal shows many transitions, some of which can be assigned to $Y3$, supporting the presence of 90-degree exchange coupled excitons inside the $Y3$ state. Finally, the level diagram for three interacting triplets is more complicated than Fig.~\ref{figmmODMRy}.b. In addition to several lower spin states it contains two quintet manifolds and a S=3 septet. The two quintet manifolds can explain why precisely two families of crossings were seen in \cite{bayliss2018}. An experimental signature of an S=3 state in MPL would be a longer sequence of crossings $\Delta/n$ with $n=1,2,3,4$. While we have not yet managed to identify such a sequence conclusively, we report some possible signatures in SI together with the in-phase high-field ODMR lineshapes.

In conclusion, we have studied the multi-exciton states that emerge from singlet fission under a high excitation fluency in the model fission material TIPS-Tetracene. We have shown that a transition takes place between two dominant long-lived species, the $U2$ at low light fluency and the $V2$ with exchange coupling in the $m \rm Tesla$ range dominating at high fluency. These weakly coupled states display an intricate family of transitions in ODMR and level crossings in MPL, which are consistent with our theoretical predictions. $V2$ shares many photo-physical properties with a strongly interacting state giving a characteristic 5.5 Tesla dip in the magneto-fluorescence which grows at high excitation. This is evidenced by their matching PL spectra and by the behavior of their high-frequency ODMR spectra at the 5.5 Tesla level crossing. We suggest that this strongly coupled state is composed of three neighbor triplet excitons forming an overall quintet state. The series of high-field dips in PL are then explained as level crossings with the triplet manifold. We confirm this assignment through high-field ODMR experiments at the level crossing. Our research opens a path to study photo-physics in the cascade of multiexciton states in fission materials.
\\

\noindent
\textbf{Acknowledgements}
 
We thank S. Bayliss and L. Weiss for helpful discussions. This project was supported by funding from ANR-20-CE92-0041 (MARS) and IDF-DIM SIRTEQ, and the European Research Council (ERC) under the European Union's Horizon 2020 research and innovation programme (grant Ballistop agreement no. 833350). V. Derkach acknowledges the kind hospitality from CNRS Gif-sur-Yvette.

\bibliography{WQodmr.bib}

\begin{widetext}

\newpage
\section*{Supplementary Information}
\setcounter{equation}{0}
\renewcommand{\theequation}{S\arabic{equation}}
\setcounter{figure}{0}
\renewcommand\thefigure{S\arabic{figure}}
\renewcommand{\figurename}{Supplementary Figure}

\section{Supplementary ODMR spectra}

\begin{figure}[h]
\begin{tabular}{c}
\centering
  \includegraphics[clip=true,width=0.5\columnwidth]{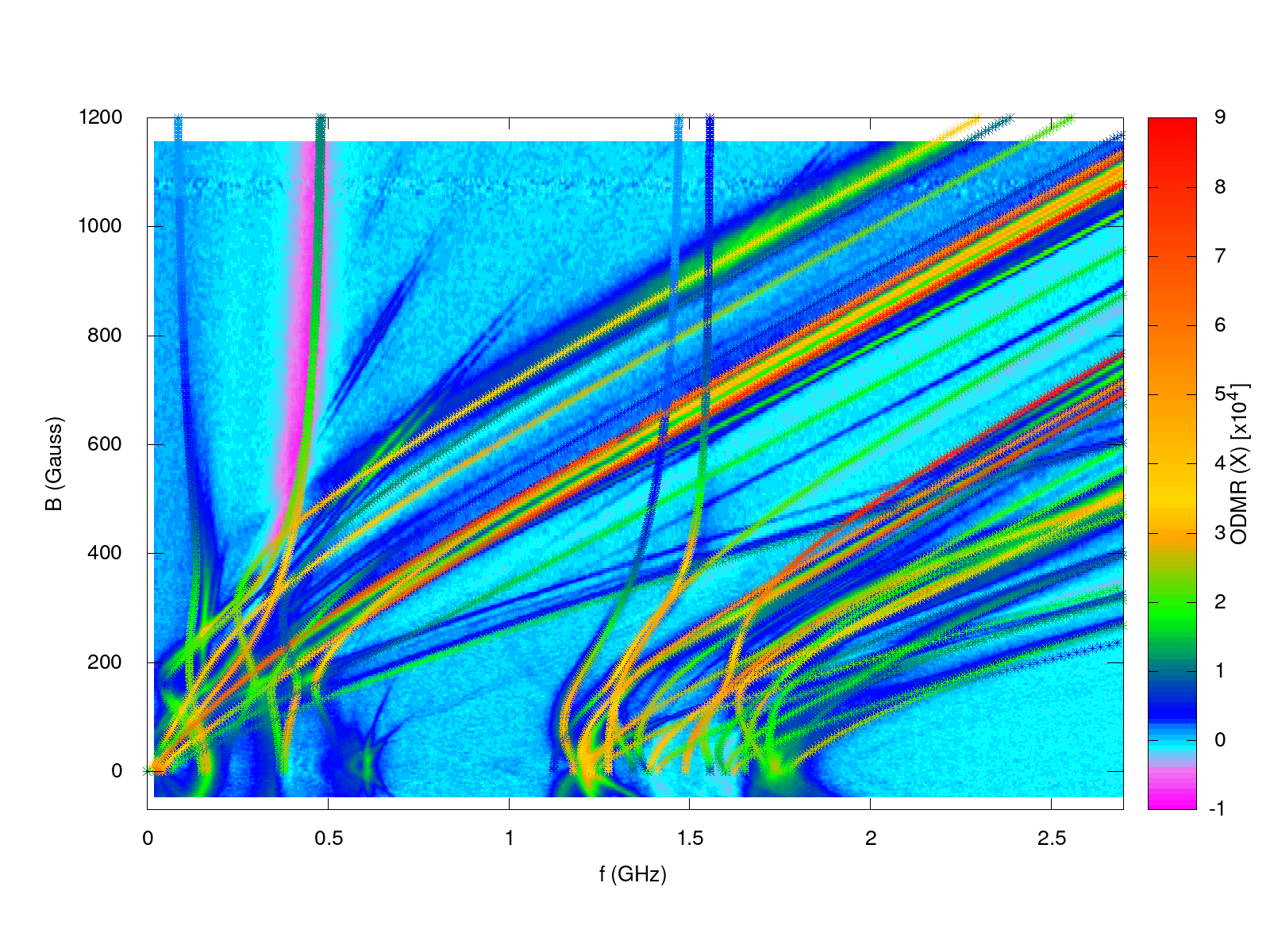}
  \includegraphics[clip=true,width=0.5\columnwidth]{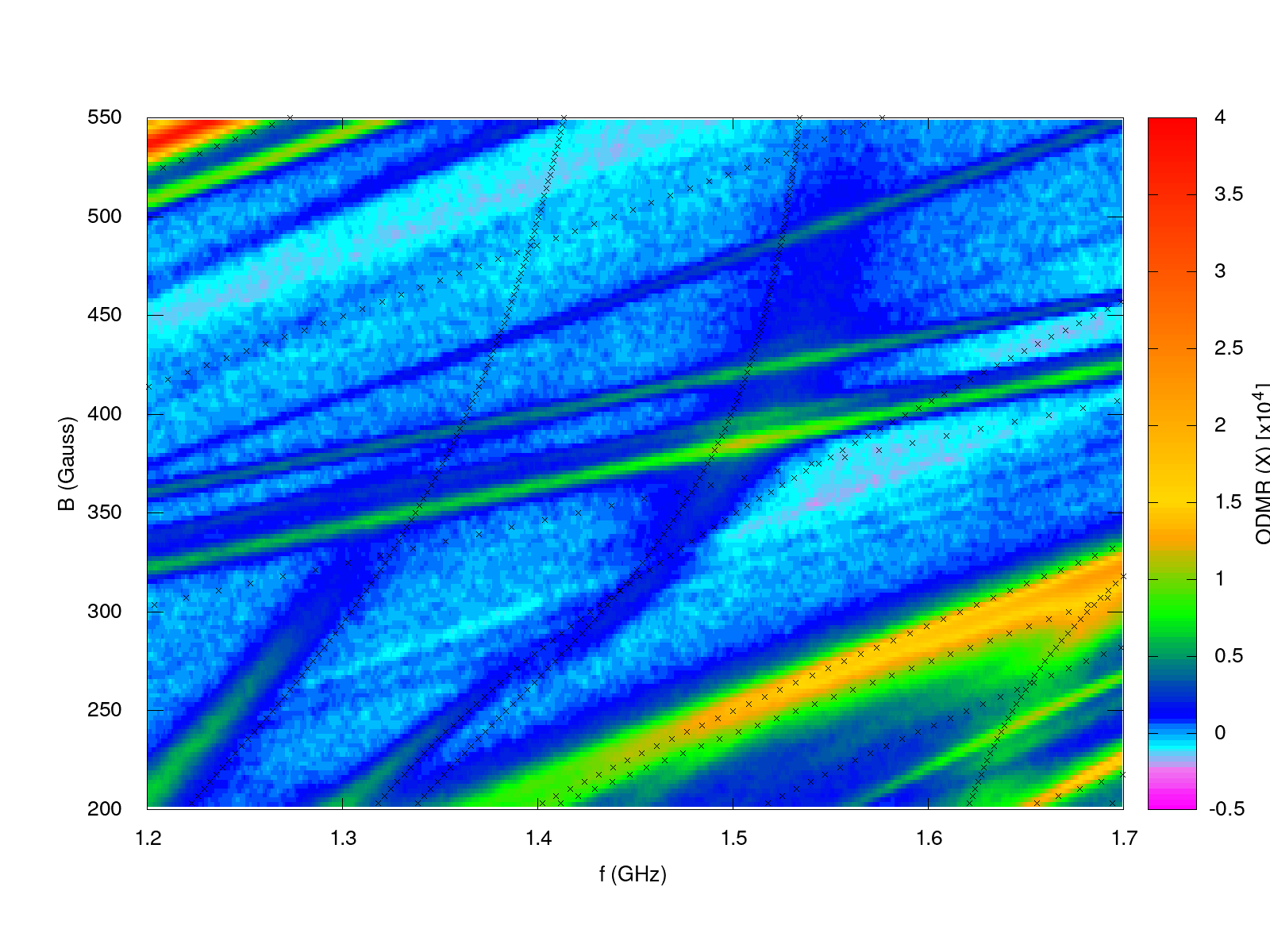} 
  \end{tabular}
\caption{ODMR pattern measured with a microwave power amplifier, which allows the microwave power to be increased from 10 dBm to 30 dBm at the expense of introducing spurious extra harmonics in the spetrum. The theoretical predictions of the transitions by weakly-coupled 90-degree stacked triplet-exciton pairs overlaid on the pattern with dotted lines, coinciding with the experimental map. The color of the doted lines indicates the dipole moment of the corresponding transitions. The enlarged map at around 1.5 GHz and 400 Gauss in the bottom panel, shows the weak transitions precisely consistent with theory. A graphical way to visualize the correspondance between the transitions and the spin-Hamiltonian eigenstates is provided in Fig.~\ref{figsmapamp2}.
}
\label{figsmapamp}
\end{figure}

\newpage
\begin{figure}[H]
\begin{tabular}{c}
\centering
  \includegraphics[clip=true,width=\columnwidth]{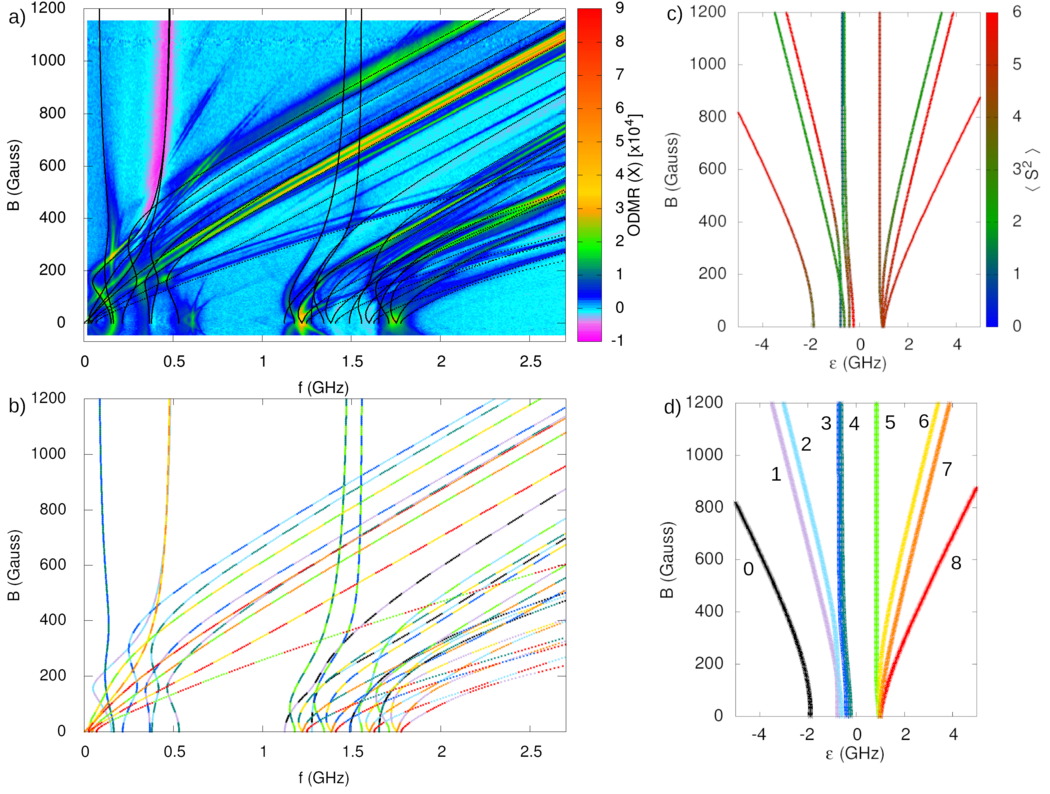} 
  \end{tabular}
\caption{Color codding of all the visible transitions to enable their identification from the spin Hamiltonian as well as the corresponding spin of each eigenstate. a) Shows the same data as in Fig.~\ref{figsmapamp}, with the theoretical tranistions shown as black doted lines. b) Shows only the theretical transitions in color-dashed lines, the two colors allow to identify the eigenstates involved in the transition. The color coding for the eigenstates is given in panel d) (ground is black, first excited state is violet ...) while the panel c) gives the total spin of the eigenstates. This coding allows to quickly find the properties of each visible transition. For example we can find a visible transition within the triplet manifold between states  $|T_0\rangle = |\epsilon_4\rangle$ (dark green) and $|T_1\rangle = |\epsilon_6\rangle$ (yellow), we can also identify the lowest frequency vertical transition $|S\rangle = |\epsilon_3\rangle$ (blue) and   $|T_0\rangle = |\epsilon_4\rangle$. 
}
\label{figsmapamp2}
\end{figure}

\begin{figure}[H]
\begin{tabular}{c}
\centering
  \includegraphics[clip=true,width=0.5 \columnwidth]{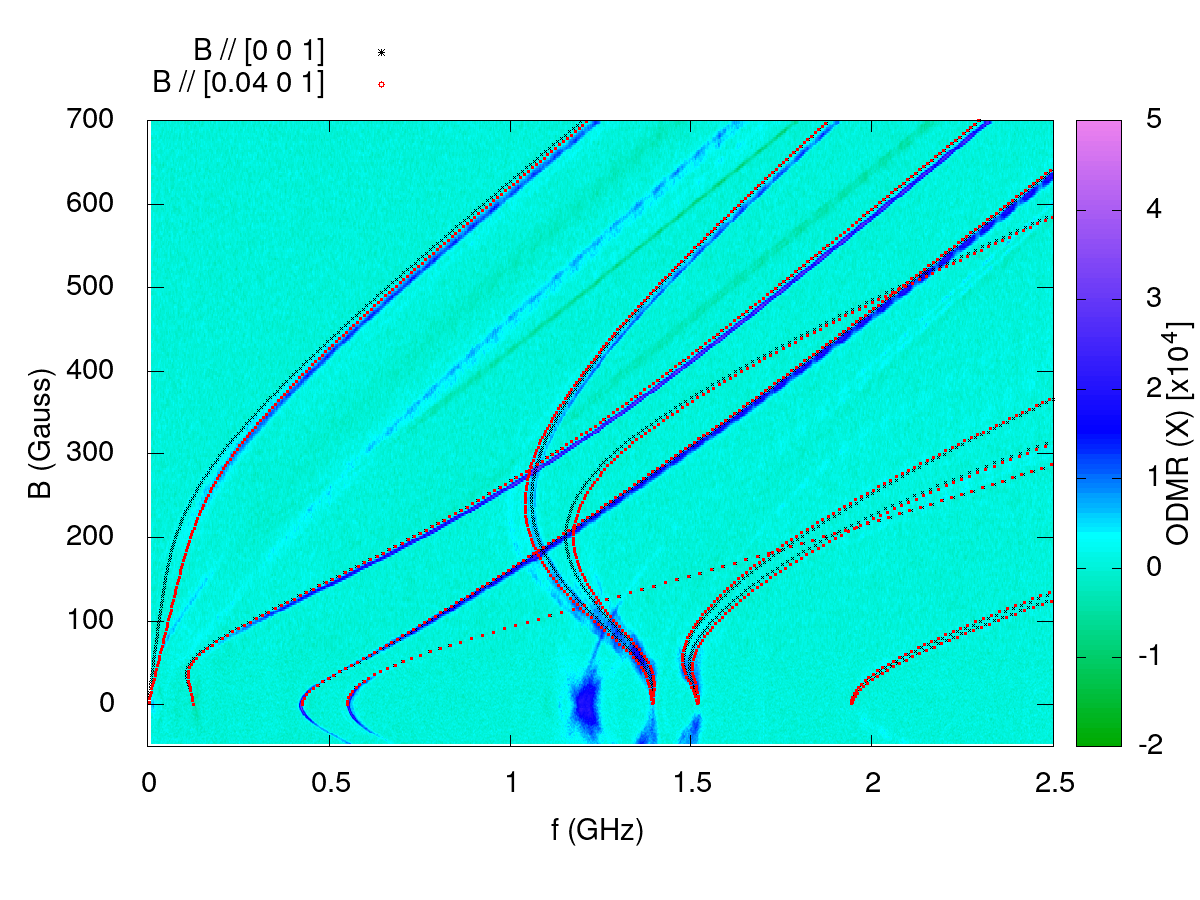}
	\includegraphics[clip=true,width=0.5 \columnwidth]{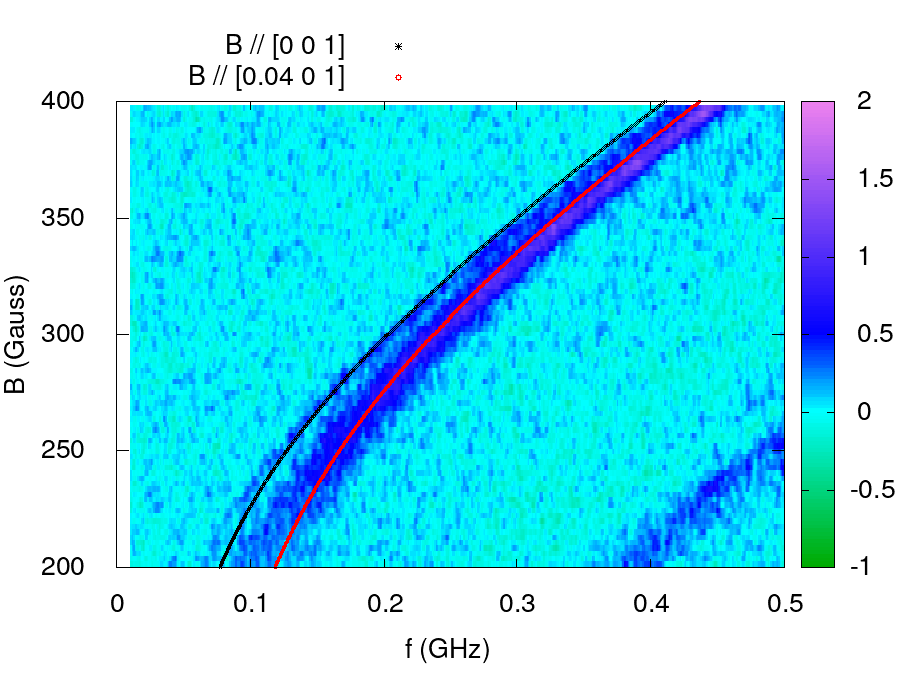} 
  \end{tabular}
\caption{Low power light excitation power Zeeman-fan diagram for the sample from Fig.~\ref{figmap} showing mainly the U2 transitions which can be used to x the orientation of the magnetic with respect to the sample more precisely. Crystal alignment suggests that B is oriented along the c axis, precise fitting of the transition lines (all the B/f range on the left panel and zoom on the right panel) allows us to x a more precise $B\; ||\; [0:04; 0; 1]$ direction. 
}
\label{figodmrpow}
\end{figure}

\begin{figure}[h]
\begin{tabular}{c}
\centering
  \includegraphics[clip=true,width=0.7 \columnwidth]{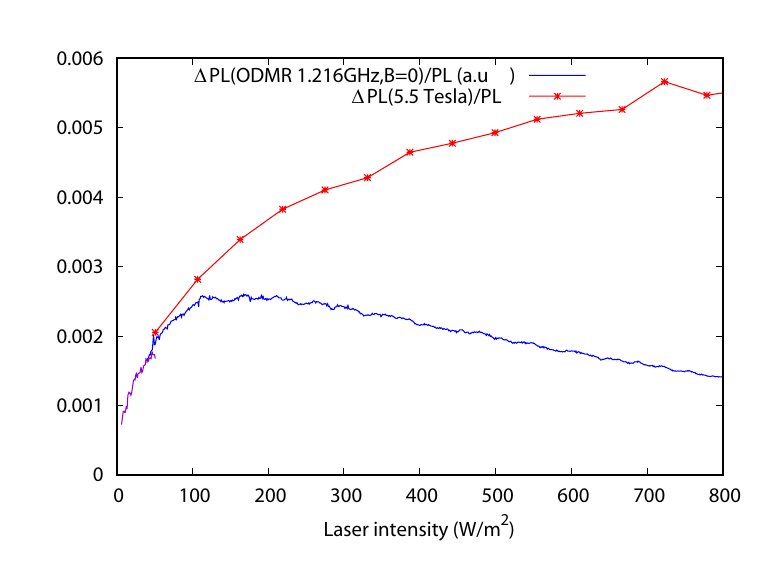} 
  \end{tabular}
\caption{Comparison of the laser power dependence of the ODMR signal from V2 at 1.2 GHz and the MPL dip from Y3 at 5.5 Tesla. We see that both signals drop at low laser intensity but the signal from V2 decreases at high laser intensity while the signal from Y3 saturates. This behavior is consistent with the identication of Y3 as a tree exciton state which can be formed as photoexcited U2 at low laser intensty creating V2 and then as a photoexcited V2 state at high laser intensity. 
}
\label{figodmrpow}
\end{figure}

\section{Mganeto-photoluminescence model}

For the model of magneto-photoluminescence at low light intensity, we try to keep the arguments as simple as possible giving up on the exact theoretical analysis of the possible mixing inside the 9 states of a triplet pair. We thus use an approximate two-level model to describe MPL. We assume that the geminate singlet pair sometimes hops to a site with low-exchange energy, which could for example be $V2$ but has enough energy to come back to its hot $\pi$ stacked configuration. Indeed triplet hopping involves molecular rearrangements which are likely to bind a triplet to the molecular site it occupies at low temperatures. This is probably what allows us to see well defined molecular sites in ODMR with lifetimes of around $100\;{\rm \mu s}$. During its lifetime in the weakly coupled state there will be a mixing between singlet and the nearest triplet states due to the fine-structure tensor while the magnetic field will tend to split the energy levels inducing a gap. For example in a model where we focus on the mixing between $|S\rangle$ and $|T_1\rangle$ states of a weakly coupled triplet pair, the effective spin Hamiltonian will be: 
\begin{align}
{\hat H}_2  = \frac{1}{2}\left(
\begin{array}{cc}
B_z & D_{eff} \\
D_{eff} & -B_z
\end{array}
\right)
\end{align} 
where $D_{eff}$ will describe the mixing between $|S\rangle$ and $|T_1\rangle$ and $B_z$ is the Zeeman level splitting. 

The probability of a transition from $|S>$ to  $|T_1\rangle$  during the lifetime $\tau$ of the weakly coupled state will be: 
\begin{align}
P_T = \frac{1}{\tau} \int_0^\infty e^{-t/\tau} \left| \langle T_1 | e^{i {\hat H}_2 t} | S \rangle  \right|^2 dt = \frac{1}{2} \frac{D_{eff}^2 \tau^2}{1+(B_z^2 + D_{eff}^2) \tau^2 } 
\end{align}
The produces a Lorentzian-line shape as function of $B_z$, with a squared width $\tau^{-2} + D_{eff}^2$. The expected values of $D_{eff} < D \simeq 40\;{mTesla}$ are too small to explain the experimental linewidth which is in the Tesla range, we thus have to assume that the linewidth is instead fixed by time-scale $\tau \ll D^{-1}$. However in this regime the magnetic field-induced change in $P_T \simeq D_{eff}^2 \tau^2$ will be too small compared to experiment. We can resolve this contradiction by assuming that multiple jumps to the weakly coupled state are made during the lifetime of the photo-excited singlet increasing the MPL contrast.  If the pair is in its singlet configuration it comes back to its $\pi$ stacked state and can then jump back to the weakly coupled configuration later. We estimate the increase in the total probability $P_T$ as $\tau^*/\tau$: the ratio between the life time of photo-excited singlet $\tau^*$ and the lifetime $\tau$ in the weakly coupled state, of course this is a bold approximation but we are simply trying to derive the most simple model to reproduce MPL. 
Finally we see that the Lorentzian-lineshape is not such a good fit to our data, because the experimental lineshape has much broader tails. 
A broader tails can be modeled by assuming a distribution of lifetimes $\tau$ which we assume to also be Lorentzian. This gives the following expression for the total probability of jumping to the dark state: 
\begin{align}
\langle P_T \rangle = \frac{2}{\pi} \int_0^\infty \frac{\gamma}{1 + \gamma^2 \tau^2} \frac{\tau^*}{\tau} P_T d \tau =  \frac{D_{eff}^2 \tau^*}{2 \pi \gamma} f( \frac{B_z^2 + D_{eff}^2}{\gamma^2}  ) 
\end{align}
where the function $f(x)$ is given by:
\begin{align}
f(x) = \frac{\log \; x}{1-x} 
\end{align}
For the plot in the main text we assumed $D_{eff} = 20$mTesla, $\gamma^{-1} = 3\;{\rm ps}$ and $\tau^{*} = 200 \;{\rm ps}$, this is a reasonable time scales for a hot singlet photo-excitation generated by singlet fission. We note that this suggests that there is a complex hot dynamics after singlet fission where the pair of triplet-excitons explores multiple sites. This expression can fit only the low magnetic field part of the MPL, to describe the MPL at high magnetic field we can instead consider a $\Delta g$ mechanism which will induce some mixing between $|S\rangle$ and  $|T_0\rangle$. In this case the effective Hamiltonian will instead be: 
\begin{align}
{\hat H}_2'  = \frac{1}{2}\left(
\begin{array}{cc}
J & B_z \Delta g  \\
B_z \Delta g  & -J
\end{array}
\right)
\end{align}
where the J represents exchange coupling in the weakly-coupled state and $\Delta_g$ is the differential Lande factor.
The same reasoning as before can be applied to ${\hat H}_2'$ leading to:
\begin{align}
\langle P_T' \rangle = \frac{(B_z \Delta g)^2 \tau^*}{2 \pi \gamma} f( \frac{J^2 + (B_z \Delta g)^2}{\gamma^2}  ) 
\end{align}
Since the probabilities $\langle P_T \rangle$ and $\langle P_T' \rangle$ are small and correspond to different loss channels we approximate:
\begin{align}
-\frac{\Delta PL}{PL} = \langle P_T \rangle + \langle P_T' \rangle
\end{align}
where $-\Delta PL/PL$ is the relative loss of PL due to magnetic field effects counted from the maximal PL as function of magnetic field. 
This is the formula that is used for the fits shown in the main figure, the high field $\langle P_T' \rangle$ part essentially gives a parabolic decrease with little reliable information on the underlying parameters ($\tau^*$ a $\gamma$ where fixed to the values obtained from the zero field fit).

\section{Singlet-triplet mixing in a triplet-pair model}

Here we comment on the physical origin of the singlet-triplet mixing parameterized by the matrix element $D_{eff}$ in the previous section. It turns out that fine structure of the individual triplets won't induce any mixing between  $|T_1\rangle$ and singlet  $|S\rangle$, for any zero magnetic field Hamiltonian containing fine structure terms, dipole-dipole interactions and exchange energy ${\hat H}_V$ we find the selection rule $\langle T_{1,0,-1}| {\hat H}_V | S \rangle = 0$.

The dipole-dipole and exchange interactions are invariant under the exchange of the two triplets, thus the absence of matrix elements between singlet wavefunctions (symmetric under triplet exchange) and triplet (anti-symmetric) is not surprising.
However this argument does not apply to the fine-structure terms which can be different for the two-spins, additional symmetries are need to explain this result. The Hamiltonian ${\hat H}_V = \sum {\hat H}_n$ is the sum of terms which transform as a scalar under rotations. If for each-of the terms ${\hat H}_n$ we can find a reference frame where we can show $\langle T_{1,0,-1}| {\hat H}_n | S \rangle = 0$ this will then imply that $\langle T_{1,0,-1}| {\hat H}_V | S \rangle = 0$ in all frames. The fine structure terms can be reduced to ${\hat S}_z^2$ by choosing the $z$ axis aligned along the molecular basis directions. We are thus left with three matrix elements to consider $\langle T_{1,0,-1}| S_z^2 | S \rangle$. 

Time reversal symmetry enforces two additional vanishing matrix elements. To see this we need to consider two groups of states $|\psi\rangle$ and $|\psi'\rangle$. The states $|\psi\rangle = |S\rangle, |Q_0\rangle, \frac{|Q_2\rangle+|Q_{-2}\rangle}{\sqrt{2}}, \frac{|Q_1\rangle-|Q_{-1}\rangle}{\sqrt{2}}, \frac{|T_1\rangle-|T_{-1}\rangle}{\sqrt{2}}$ remain unchanged by time-reversal symmetry (in the two-triplet basis time reversal symmetry operation is given by $9\times9$ matrix with alternating $1, -1$ on the anti-diagonal). The other states $|\psi'\rangle = |T_0\rangle,  \frac{|T_1\rangle+|T_{-1}\rangle}{\sqrt{2}}, \frac{|Q_1\rangle+|Q_{-1}\rangle}{\sqrt{2}}, \frac{|Q_2\rangle-|Q_{-2}\rangle}{\sqrt{2}}$ instead acquire a $-$ sign. We can show that for any time-reversal symmetric Hamiltonian ${\hat H}$ the matrix elements between states of different symmetry $\rangle \psi| {\hat H} | \psi' \langle$ are necessarily pure imaginary complex numbers.

Coming back to the fine-structure matrix elements we notice that in the tensor product triplet basis ${\hat S}_z^2$ is a real matrix as all the components of $| T_0 \rangle, |T_1 + T_{-1}\rangle, |S\rangle$. We can thus conclude $\langle S| {\hat S}^2_z | T_0 \rangle = \langle S| {\hat S}^2_z | T_1 + T_{-1} \rangle = 0$ based on time reversal symmetry. This leaves a last matrix-element  $\langle S| {\hat S}^2_z | T_{1} - T_{-1} \rangle $, it can be shown to be zero by direct calculation, or we can use the invariance of ${\hat S}^2_z$ under reflections on the $x,y$ plane to transform $T_{1} - T_{-1} \rightarrow T_{1} + T_{-1}$. 

Having completed this analysis we can think about the physical ingredients which can lead to mixing between singlet and triplet at zero magnetic field. First we can see that if we include spin-orbit interactions in the problem the orbital momentum must also be included in the time-reversal symmetry argument. For example assuming $p$ orbitals we can check that $|S, p_z\rangle$ and $|T_0, p_x\rangle$ will behave in the same way under-time reversal symmetry reducing the number of constraints enforcing the selection rule $\langle T_{1,0,-1}| {\hat H} | S \rangle = 0$ . Inversion symmetry and the related symmetry under exchange of the two triplets also constrained several matrix elements. The Dzyaloshinskii-Moriya interaction (DMI) exactly matches these two constraints, it is spin-orbit induced and inversion symmetry breaking. It has the form:
\begin{align}
{\hat H}_{DMI} = \mathbf{D}_{12} \cdot (\hat{\mathbf{S}}_1 \times \hat{\mathbf{S}}_2) 
\end{align}
where $\mathbf{D}_{12}$ is a vector. 
We can check that DMI indeed provides non-zero matrix elements $\langle T_{1,0,-1}| {\hat H}_{DMI} | S\rangle$. Since DMI is anti-symmetric under the exchange of the two triplets, its matrix elements inside triplet and quintet manifold identically vanish $\langle T_n| {\hat H}_{DMI} | T_m \rangle = \langle Q_n| {\hat H}_{DMI} | Q_m \rangle = 0$. Thus a non-zero DMI is interaction won't affect the $\pi$-stacked quintet Hamiltonian which we measured quite precisely, this may explain why we didn't have to include this term so for in our spin-Hamiltonian.

In conclusion, DMI interactions seem to provide the main mechanism for singlet to triplet mixing for exciton pairs at zero magnetic field. The parameter $D_{eff} = 20\;{\rm mTesla}$ from the previous section, can be physically interpreted as the strength of the DMI interaction between the two triplets.

\section{High magnetic field ODMR} 

\begin{figure}[h]
\includegraphics[clip=true,width=0.8\columnwidth]{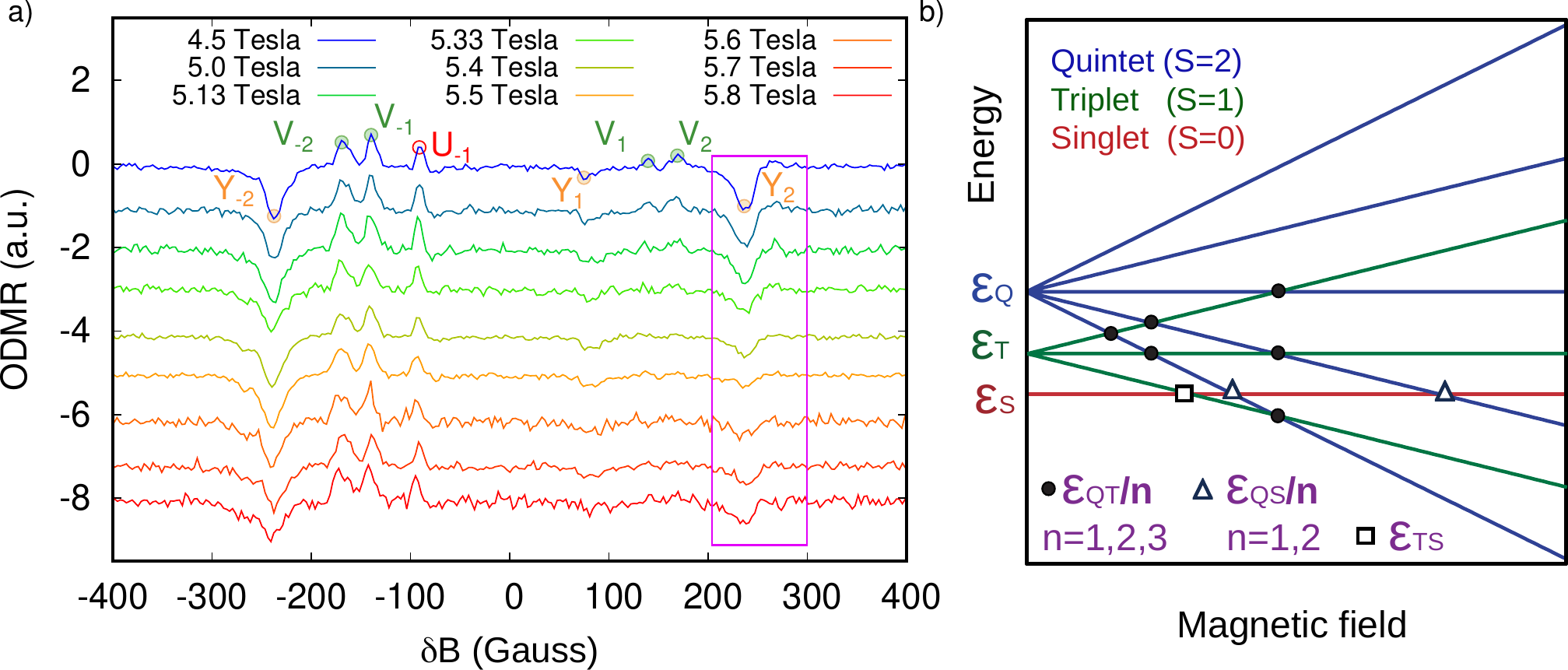}  
\caption{High field ODMR spectra at around 5.5 T level crossing. Under high field, we clearly identify all the quintet-like transitions from $V2$ ($V_{2,1,-1,-2}$) while at most only one resonate peak from $U2$ is visible  ($Q_{-1}$). A new group of negative transition from Y3 ($Y_{-2,1,2}$) appears under high magnetic field. The $Y_{2}$ transition is strongly suppressed at the 5.5T level crossing, suggesting that the $Y3$ transitions can be attributed to this strongly coupled state. b) Energy diagram of two triplet pairs, where black dots label the level crossing between quintet and triplet with the expected sequence in magnetic field of $\epsilon_{QT}/1$, $\epsilon_{QT}/2$ and $\epsilon_{QT}/3$, where $\epsilon_{QT}$ is the energy splitting between triplet and quintet states at zero field. The square and triangle label the crossing between quintet to single and triplet to single, respectively. The suggested position of the 5.5T level crossing allows us to explain the suppression of the $Y_2$ transition since at this crossing both the $|+2>$ and $|+1>$ quintet states are depleted due to the crossing with the triplet manifold. The level crossing diagram for coupled states of three excitons will look similar but with a longer series of crossings.
}
\label{figmmODMRx}
\end{figure}

\begin{figure}[h]
\includegraphics[clip=true,width=0.8\columnwidth]{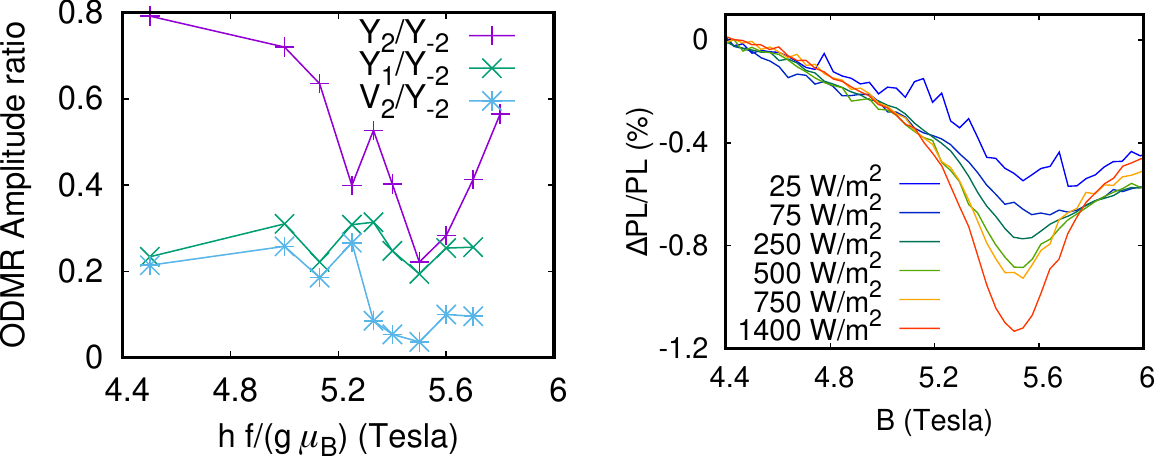}  
\caption{The ratio $Y_2/Y_{-2}$ as a function of the magnetic field, in which a dip at 5.5 T matches the level crossing lineshapes at 5.5 in magneto photoluminescence. While for $Y_1/Y_{-2}$, this correlation does not appear within the noise level. 
}
\label{figszoom}
\end{figure}

To link magnetic resonance and high-field level crossing experiments we are looking for transitions whose intensity changes at the level crossing of $5.5$ Tesla. The exciting MW electromagnetic field from the waveguide depends on the frequency because of reflection at the end of the waveguide and from the sample oscillating with frequency. We estimate that the ODMR spectra at nearby maxima should correspond to similar excitation power. This is the case for traces at 4.5 Tesla and 5.4 Tesla on Fig.~\ref{figmmODMRx}a, which can be compared directly without rescaling. This allows us to see that the amplitude of peaks  $Y_{-2}, V_{-2}$ and $V_{-1}$ remain unchanged at the $5.5$ level crossing. We then used this family of peaks to rescale the curves at other excitation frequencies keeping the amplitude of these transitions constant leading to the data in Fig.~\ref{figmmODMRx}.a). A clear dip of the $Y_{-2}$ transition appears at $5.5$ Tesla (the ratio $Y_{2}/Y_{-2}$ as a function of the magnetic field is compared to MPL S.I.). In \cite{bayliss2018} the 5.5 level crossing was associated with a level crossing between the $|-2\rangle$ of the quintet and singlet manifolds. For a level crossing with the singlet manifold, we could expect an enhancement of the $Y_2$ transition associated with the depopulation of the geminate singlet which manifests as a drop of PL at the level crossing.  Our data instead suggests a drop of the  $|-2\rangle$ population consistent with the level crossing with a dark short-lived state. This points to a level crossing with the triplet manifold and a non-geminate character of the level crossing at 5.5 Tesla consistent with the power dependence on Fig.~\ref{figMPLandSpec}. The level crossing diagram for a pair of triplet excitons is shown in Fig.~\ref{figmmODMRy}b. There are three crossings between quintet and triplet. The crossings form a sequence $\Delta/n$ where $\Delta$ is the zero-field splitting between quintet and triplet manifolds and $n$ numbering the crossings.

Fig.~\ref{figmmODMRx}a shows the ODMR spectra around the magnetic field of the MPL dip.
From the visible transitions, some belong to species that we already identified.
The weakly coupled quintet transitions marked as $V_{-2,-1,1,2}$ correspond to transitions inside the quintet manifold that are shown in Fig.~\ref{figmap}. The index refers to states involved in the transition (for example $V_{-2}$ is between $|-2\rangle$ and $|-1\rangle$ quintet states). Consistent with the trend of $U2$ becoming less visible at high magnetic fields, only a single transition $Q_{-1}$ can be assigned to this family. Due to the switch from the strip-line ($f < 20\;{\rm GHz}$) to waveguide excitation $f > 120\;{\rm GHz}$) we cannot continuously follow the evolution of the ODMR spectrum from low to high magnetic fields. This introduces a challenge to the precise identification of ODMR lines and we cannot exclude at high magnetic field overlap between $Q_{-1}$ and the dense transitions from the W state. In addition to the positive ODMR peaks from the $V$ and $U$ families, a series of negative peaks are also visible, which we classified into the $Y3$ group. This family of transitions did not give rise to a strong signal at low magnetic fields.

The position of the $Y3$ family peaks seems to match the expected positions for a strongly bound 90-degree exciton on the nearest 90-degree sites. An explanation could be that the 5.5 Tesla level crossing is a coupled state between a triplet exciton and a triplet pair on the nearest 90-degree tilted neighbor. We notice that a crossing between two manifolds with spin $S$ and $S'$ will follow the sequence: $\Delta, \Delta/2, ..., \Delta/(S + S')$. Thus a coupled state of three triplet excitons is likely to manifest as longer sequences of $\Delta/n$ level crossings in the MPL data. 
We performed very precise MPL experiments with a modulation coil looking for evidence of additional level crossings, finding possible evidence of a continuation of the $\Delta/n$ sequence beyond three terms. 
The existence of $S=3$ states with 3 interacting triplets would also explain the presence of two families of level crossings in the experiments of \cite{bayliss2018} with different exchange energies $5.5$ and $4.4$ Tesla. The two families could then correspond to crossings between $S = 3$ and $S = 1$ manifolds (4 crossings) and between the $S = 3$ and $S = 2$ manifolds (5 crossings) of the ensemble of 3 excitons, as manifested in Fig.~\ref{figstableQT}a.

\section{Extra transitions in MPL}

\begin{figure}[h]
\includegraphics[clip=true,width=0.6\columnwidth]{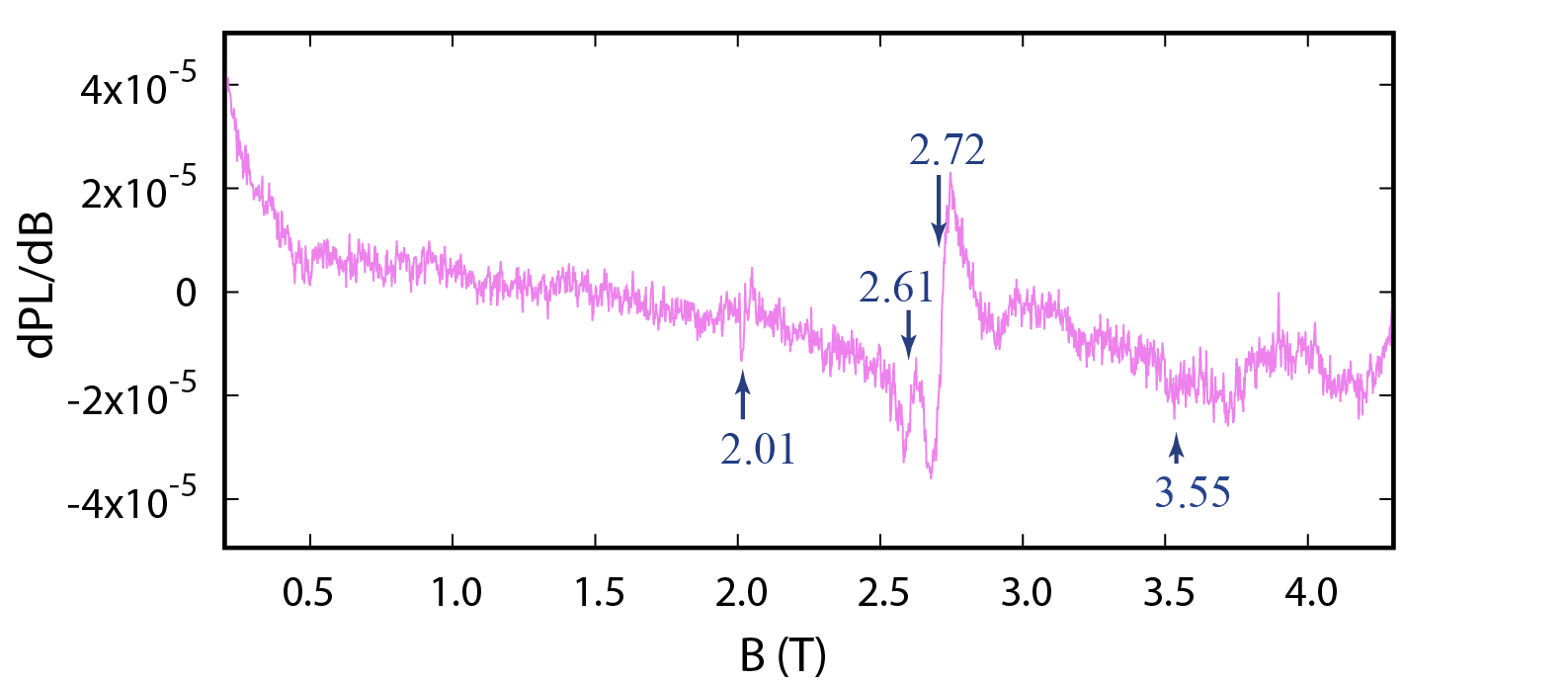}  
\caption{Differential Magneto photoluminescence measured using an additional field modulation coil allowing a lock-in detection of the integrated-PL at the field-modulation frequency. Except for the level crossing peak at around 4.8 T, a weaker transition feature at 2.1 T is also observable, which could be associated with Y transitions at low-field. These presence of MPL peaks at lower magnetic field can be understood from the level-crossing series from septet to quintets (see Figs.~\ref{figstableQT},\ref{figstableST}). 
}
\label{figsmpl}
\end{figure}

\begin{figure}[h]

\includegraphics[clip=true,width=0.4\columnwidth]{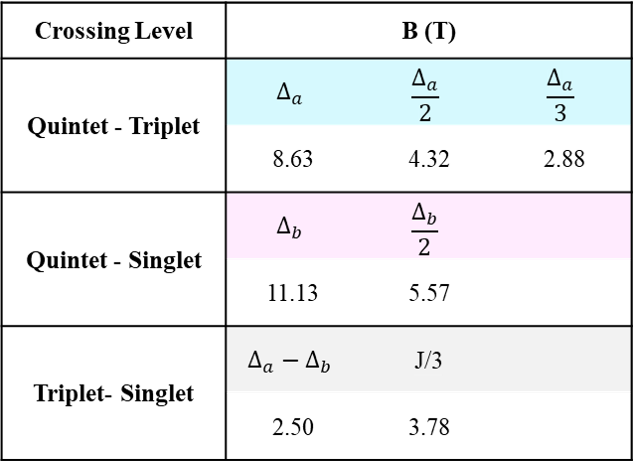}  
\caption{The expected energy sequence for level crossing between two spin manifolds for two triplet pairs, attempting to explain the presence of two families of level crossings at $11.1$ Tesla$/n$ and  $8.6$ Tesla$/n$. This leads to a quintet-singlet level crossing at 2.5 Tesla instead of 3.78 Tesla in the experiment. The 3.78 Tesla is consistent with a level crossing of a quintet state with triplet and singlet states in a model of two exchange-coupled triplets as repoted in \cite{bayliss2018} for which a $8.6$ Tesla$/n$ ($n=1,2,3$) sequence is also formed. However, this sequence no longer holds if the position of the singlet, triplet and quintet energy is not described by a simple exchange model.
}
\label{figstableQT}
\end{figure}

\begin{figure}[h]

\includegraphics[clip=true,width=0.8\columnwidth]{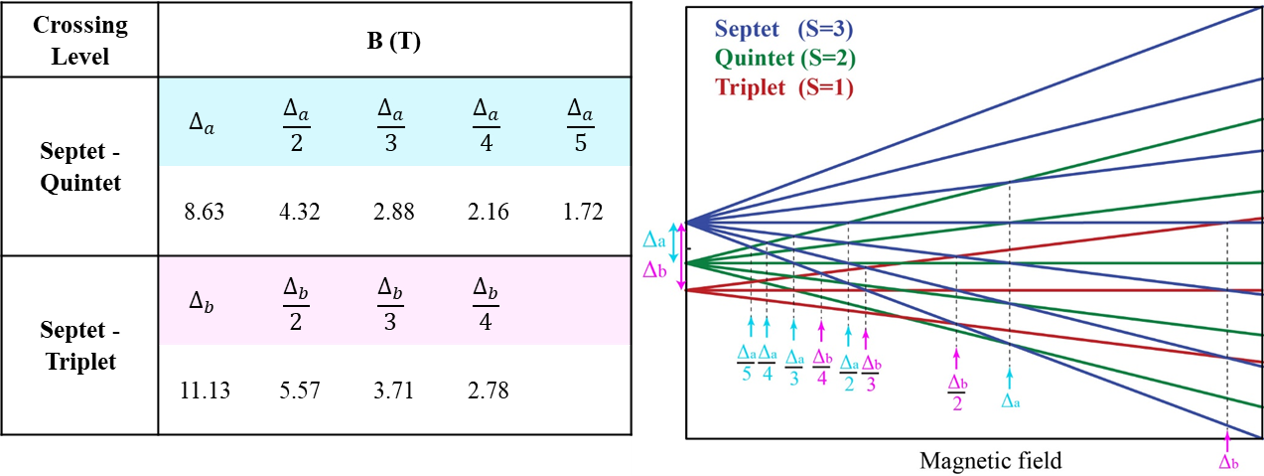}  
\caption{Simple senario of the energy levels of three interacting triplets modeled as exchange coupled S=2 and S=1 states. With the highest spin order of S=3, the level crossing field could be extended to a lower field, e.g. a resonance peak at  $\Delta_{\alpha}/4$ (2.16 T) is expected belonging to the crossing between septet and quintet, which can match with the weak resonant peak in Fig.~\ref{figsmpl}. For now we do not consider this as sufficient evidence for a $S=3$ state and therefore we have only showed the simplest possible energy diagram in Fig.~\ref{figmmODMRy} which explains areadly most of the experimental features. 
}
\label{figstableST}
\end{figure}

\end{widetext}
\end{document}